\documentclass{PoS}

\title{The Integrable Bootstrap Program at Large N and Its Applications in Gauge Theory}

\ShortTitle{Planar Integrable Bootstrap}

\author{\speaker{Axel Cort\'es Cubero}\thanks{Travel funds for this conference were provided by the Doctoral Student Research Grant, Competition $\# $ 8, from the Graduate Center, CUNY. Research supported by a Dean K. Harrison award from the Graduate Center, CUNY, and a PSC CUNY research grant. Work done in collaboration with Peter Orland.}\\
        Graduate School and University Center, City University of New York, 365 Fifth Avenue, New York, NY 10016, U.S.A.\\
        Bernard Baruch College, City University of New York, 17 Lexington Avenue, New York, NY 10010, U.S.A.\\
        E-mail: \email{acortes\_cubero@gc.cuny.edu}}

\abstract{We present results for the large-$N$ limit of the (1+1)-dimensional principal chiral sigma model. This is an asymptotically-free $N\times N$ matrix-valued field with massive excitations. All the form factors and the exact correlation functions of the Noether-current operator and the energy-momentum tensor are found, from Smirnov's form-factor axioms. We consider (2+1)-dimensional $SU(\infty)$ Yang-Mills theory as an array of principal chiral models with a current-current interaction. We discuss how to use our new form factors to calculate physical quantities in this gauge theory.}

\FullConference{31st International Symposium on Lattice Field Theory - LATTICE 2013\\
		July 29 - August 3, 2013\\
		Mainz, Germany}

\begin{document}

\section{Introduction}

 An integrable field theory has an infinite number of nontrivial conservation laws. These conservation laws restrict the form of the scattering matrix and matrix elements of operators. One can often write down exact results without evaluating any Feynman diagrams. This approach is called the integrable bootstrap program \cite{smirnov}, \cite{babujian}.
 
Theories of a matrix-valued field (such as Yang-Mills theory) in the large-$N$ limit have not been completely solved until now. The large-$N$ limit of matrix-valued field theories is much more interesting than that of isovector models, since the Feynman diagrams are planar, and the Green's functions are not trivial. A modern review of large-$N$ methods in gauge theories can be found in Reference \cite{panero}. 

Theories of isovector-valued fields (such as the $O(N)$ sigma model and the $SU(N)$ Chiral Gross-Neveu model) have been solved in the large-$N$ limit \cite{polyakov}. In this case the Feynman diagrams become linear, rather than planar. Matrix elements of local operators have been found independently in the integrable bootstrap program in these theories. Full agreement is found between the integrable bootstrap and the large-$N$ expansion \cite{babujian}. 
%The integrable bootstrap has yielded successful results in (1+1)-dimensional field theory. In theories of isovector-valued fields (such as the $O(N)$ sigma model, and the $SU(N)$ Chiral Gross Neveu model), the results can later be compared against the $1/N$ expansion or perturbation theory. These theories can be solved independently at large $N$ because their Feynman diagrams become linear (rather than planar) \cite{polyakov}.

We have made significant progress in solving the principal chiral sigma model using the integrable bootstrap and the large-$N$ limit simultaneously \cite{correlations}, \cite{multiparticle}, \cite{fieldformfactors}.  

The principal chiral sigma model is the (1+1)-dimensional theory with the action
\begin{eqnarray}
S=\frac{N}{2g_0}\int\,d^2x\, {\rm Tr}\,\,\partial_\mu U^\dag(x)\partial^\mu U(x),\nonumber
\end{eqnarray}
where $U(x)\in SU(N)$. This action has a global $SU(N)\times SU(N)$ symmetry, given by $U(x)\to V^L U(x) V^R$, where $V^{L,R}\in SU(N)$. The Noether currents associated with these symmetries are
\begin{eqnarray}
j_\mu^L(x)_a^c=\frac{-iN}{2g^2}\partial_\mu U_{ab}(x)U(x)^{\dag\,bc}(x),\,\,\,\,\,\,\,j_\mu^R(x)^d_b=\frac{-iN}{2g^2}U^{\dag\,da}(x)\partial_\mu U_{ab}(x),\label{noether}
\end{eqnarray}
respectively, where we have included the color indices, $a,b,c,d=1,\dots, N$, explicitly.

The principal chiral model is asymptotically free, and its excitations are massive color dipoles. This model is integrable, and this property has been used to find the two-particle S-matrix \cite{wiegmann}:
\begin{eqnarray}
&&\left._{\rm out}\langle P,\theta_1',c_1,d_1;P,\theta_2',c_2,d_2|P,\theta_1,a_1,b_1;P,\theta_2,a_2,b_2\rangle\right._{\rm in}=S(\theta,N)^{c_1, d_1; c_2, d_2}_{a_2, b_2; a_1, b_1}\langle \theta_1^\prime\vert \theta_1\rangle \langle \theta_2^\prime\vert \theta_2\rangle \nonumber\\
&&\,\,\,\,\,\,\,\,\,\,\,\,\,\,\,\,\,\,\,\,\,=S(\theta,N)\left(\delta_{a_1}^{c_1}\delta_{a_2}^{c_2}-\frac{2\pi i}{N\theta}\delta_{a_1}^{c_2}\delta_{a_2}^{c_1}\right)\times\left(\delta_{b_1}^{d_1}\delta_{b_2}^{d_2}-\frac{2\pi i}{N\theta}\delta_{b_1}^{d_2}\delta_{b_2}^{d_2}\right)\langle \theta_1'|\theta_1\rangle\langle \theta_2'|\theta_2\rangle,\label{smatrix}
\end{eqnarray}
where $P$ labels a particle, $A$ labels an antiparticle and $\theta_i$ is the rapidity of the $i$-th particle, defined by the parametrization of energy and momentum: $E_i=m\cosh\theta_i,\,\,p_i=m\sinh\theta_i$, and $\theta=\theta_1-\theta_2$.  The $i$-th particle has a left-color index $a_i$ and a right color index $b_i$. In the large-$N$ limit, the function $S(\theta,N)$ is simply $S(\theta,N)=1+\mathcal{O}\left(\frac{1}{N^2}\right)$ \cite{wiegmann}. The terms in (\ref{smatrix}) that are suppressed by $1/N$ cannot be discarded if the incoming particles share a contracted color index. In this case the delta functions in (\ref{smatrix}) produce a new factor of $N$. The S-matrix for particle-antiparticle scattering can be found from (\ref{smatrix}) using crossing symmetry.

Some interesting calculable physical quantities in an integrable theory are the form factors of a local operator $\mathcal{O}$. Form factors are defined as the matrix elements of this operator between a state of incoming excitations and the vacuum. Explicitly:
\begin{eqnarray}
F^{\mathcal{O}}=\langle 0\vert \mathcal{O}(x) \vert {\rm state\,with\,incoming\,particles\,and\,antiparticles}\rangle.\label{incomingstate}
\end{eqnarray}
If all the form factors are known, two-point correlation functions can be calculated by inserting a complete set of intermediate states:
\begin{eqnarray}
\langle0\vert \mathcal{O}(x)\mathcal{O}(y)\vert 0\rangle=\sum_{\Psi}\langle 0\vert \mathcal{O}(x)\vert \Psi\rangle \langle \Psi \vert \mathcal{O}(y)\vert 0\rangle.\nonumber
\end{eqnarray}

Some recent progress has been made in understanding the principal chiral model in terms of resurgence theory. In Reference \cite{unsal}, the role of non-perturbative saddle points in the path integral of the chiral model was examined. These configurations are not topologically stable, given the homotopy group $\pi_2(SU(N))=0$. It is shown how these configurations may explain the mass gap of the theory. In our approach, we assume the existence of a mass gap, without providing proof. More applications of resurgence in quantum field theory are found in References \cite{resurgence}.

An alternative large-$N$ limit of the principal chiral model, which is not the 'tHooft limit, has been examined in References \cite{kazakov}. The mass of the heaviest bound state is fixed as $N$ goes to infinity. The mass spectrum thus becomes continuous.

In what follows we find all the form factors of the Noether-current operators (\ref{noether}) and the energy-momentum tensor in the 'tHooft limit. We then write down exact expressions for the two-point functions of these operators. We mention how these results can be applied to Yang-Mills theory in 2+1 dimensions.

\section{Exact Form Factors and the Smirnov Axioms}

We will next discuss the form factors of the current operator $j_\mu^L(x)_a^c$ at large $N$. This operator has two left-handed color indices and no right-handed color index. This implies that in a matrix element of the form (\ref{incomingstate}), the incoming state must have an equal number of particles and antiparticles. The color indices of these excitations and the current must be contracted in a $SU(N)\times SU(N)$-rotation-invariant way. Using color and Lorentz invariance, we can write down an {\em Ansatz} for the current form factor with $M$ particles and $M$ antiparticles:
\begin{eqnarray}
&&\langle 0\vert j_\mu^L(x)_{a_0 a_{2M+1}}\vert A,\theta_1,b_1,a_1;\dots;A,\theta_M,b_M,a_M;P,\theta_{M+1},a_{M+1},b_{M+1};\dots;P,\theta_{2M},a_{2M},b_{2M}\rangle\nonumber\\
&&=(p_1+\cdots +p_{M}-p_{M+1}-\cdots -p_{2M})_\mu\frac{e^{-ix\cdot \sum p}}{N^{M-1}}\sum_{\sigma,\tau\in S_{M}} F_{\sigma\tau}(\theta_1,\dots,\theta_{2M})\nonumber\\
&&\times\left[\prod_{j=0}^M\delta_{a_j a_{\sigma(j)+M}}\prod_{k=1}^M\delta_{b_k b_{\tau(k)+M}}-\frac{1}{N}\delta_{a_0a_{2M+1}}\delta_{a_{l_\sigma}a_{\sigma(0)+M}}\prod_{j=1,\,j\neq l_\sigma}^M \delta_{a_j a_{\sigma(j)+M}}\prod_{k=1}^M\delta_{b_k b_{\tau(k)+M}}\right].\label{ansatz}
\end{eqnarray}
Here $\sigma$ is a permutation taking the set of numbers $0,1,2,\dots,M$ to $\sigma(0),\sigma(1),\dots,\sigma(M)$, respectively,  $\tau$ is another permutation taking the numbers $1,2,\dots,M$ to $\tau(1),\tau(2),\dots,\tau(M)$, respectively, and $l_\sigma$ is defined for a permutation $\sigma$ by $\sigma(l_\sigma)+M=2M+1$. Our expression is a sum over permutations of the color-invariant contractions of indices. The term in the the square brackets in (\ref{ansatz}) proportional to $1/N$ ensures that the current operator is traceless, {\em i.e.} $\delta_{c}^{a} j_\mu^L(x)_a^c=0$. It remains to find the function of the rapidities $F_{\sigma\tau}(\{\theta\})$, for each pair of permutations.

We can determine the functions $F_{\sigma\tau}(\{\theta\})$ at large $N$ using the Smirnov form-factor axioms \cite{smirnov}. These axioms follow from the integrability of the theory, and they place so many restrictions on the form factors that they can be found exactly, assuming maximal analyticity.  

Smirnov's scattering axiom (known also as Watson's theorem) states that the order of two excitations in a incoming state  can be switched by multiplying the form factor by a two-excitation S-matrix. Explicitly:
\begin{eqnarray}
&&\langle 0\vert j_\mu^L (x)_{a_0a_{2M+1}}\vert I_1,\theta_1,C_1;\dots;I_j,\theta_j,C_j;I_{j+1},\theta_{j+1},C_{j+1};\dots;I_{2M},\theta_{2M},C_{2M}\rangle\nonumber\\
&&=S(\theta_j-\theta_{j+1})_{C_j C_{j+1}}^{C_{j+1}^\prime C_{j}^\prime} \langle 0\vert j_\mu^L(x)_{a_0a_{2M+1}}\vert I_1,\theta_1,C_1;\dots;I_{j+1},\theta_{j+1},C_{j+1}^\prime;I_j,\theta_j,C_j^\prime;\dots;I_{2M},\theta_{2M},C_{2M}\rangle,\nonumber\\
&&\label{scattering}
\end{eqnarray}
where $I_k=A$ if the $k$-th excitation is an antiparticle and $I_k=P$ when it is a particle, and $C_k$ is the set of indices $b_k,a_k$ if $I_k=A$ and $a_k,b_k$ if $I_k=P$. The Kronecker deltas in (\ref{ansatz}) may produce new powers of $N$ when contracted with the S-matrix (\ref{smatrix}). For a given pair of permutations $\sigma,\,\tau$, the $1/N$-expanded S-matrix used to exchange two excitations is not unity only if the excitations share a contracted color index. 

Smirnov's periodicity axiom follows from crossing symmetry. This axiom states that one can replace the $2M$-th excitation in the form factor by the first excitation, shifting its rapidity by $-2\pi i$:
\begin{eqnarray}
&&\langle 0\vert j_\mu^L(x)_{a_0a_{2M+1}}\vert I_1,\theta_1,C_1;\dots;I_{2M},\theta_{2M},C_{2M}\rangle\nonumber\\
&&=\langle 0\vert j_\mu^L(x)_{a_0a_{2M+1}} \vert I_{2M},\theta_{2M}-2\pi i,C_{2M};I_1,\theta_1,C_1;\dots;I_{2M-1},\theta_{2M-1},C_{2M-1}\rangle.\nonumber
\end{eqnarray}
A similar exchange can be done for the $2M-1$-st particle, then the $2M-2$-nd and so on.

The scattering and periodicity axioms determine the function $F_{\sigma\tau}(\{\theta\})$ up to a constant. This constant is fixed by the annihilation-pole axiom. This axiom states that an incoming particle with rapidity $\theta_j$ can annihilate with an incoming antiparticle with rapidity $\theta_k$. Consequently, the function $F_{\sigma\tau}$ has a pole at $\theta_j-\theta_k=-\pi i$. The annihilation-pole axiom yields the residue of the $2M$-excitation form factor at this pole from the $2M-2$-excitation form factor. 
The constants are fixed iteratively, from the form factors with fewer particles. More details are shown in \cite{multiparticle}.

The solution obtained from the form-factor axioms is
\begin{eqnarray}
F_{\sigma\tau}(\{\theta\})=
\left\{\begin{array}{c}\frac{2\pi i(4\pi)^{M-1}}{\prod_{j=1,j\neq l_\sigma}^{M}(\theta_j-\theta_{\sigma(j)+M}+\pi i)\prod_{k=1}^{M}(\theta_k-\theta_{\tau(k)+M}+\pi i)},\,\,\,{\rm for }\,\,\sigma(j)\neq\tau(j),\,\,{\rm for\,all\,} j\\
0,\,\,\,\,\,{\rm otherwise}\end{array}\right. .\label{smirnovresult}
\end{eqnarray}
The annihilation-pole axiom ensures that the function $F_{\sigma\tau}(\{\theta\})$ vanishes for permutations where it would have double poles. 

At finite $N$, only the $M=1$ case has been solved \cite{multiparticle}. The spectrum of particles at finite $N$ includes bound states. This introduces additional  analytic structure in the functions $F_{\sigma\tau}(\{\theta\})$, which makes the problem insoluble for higher $M$, so far.

The correlation function of two current operators is computed by inserting a complete set of intermediate states:
\begin{eqnarray}
&&W_{\mu\nu}(x)_{a_0c_0e_0f_0}=\frac{1}{N}\langle0|j_\mu^L(x)_{a_0c_0} \,j_\nu^L(0)_{e_0f_0}|0\rangle\nonumber\\
&&=\sum_{M=1}^\infty \frac{1}{N!(M!)^2}\int\frac{d\theta_1\dots d\theta_{2M}}{(2\pi)^{2M}}\,e^{-ix\cdot\sum p}\nonumber\\
&&\times [\langle 0\vert j_\mu^L(0)_{a_0c_0}\vert A,\theta_1,b_1,a_1;\dots; A,\theta_M,b_M,a_M; P,\theta_{M+1},\theta_{M+1},a_{M+1},b_{M+1};\dots; P,\theta_{2M},a_{2M},b_{2M}\rangle ]\nonumber\\
&&\times [\langle 0\vert j_\nu^L(0)_{e_0f_0}\vert A,\theta_1,b_1,a_1;\dots; A,\theta_M,b_M,a_M; P,\theta_{M+1},\theta_{M+1},a_{M+1},b_{M+1};\dots; P,\theta_{2M},a_{2M},b_{2M}\rangle ]^*.\nonumber\\
&&\label{completeset}
\end{eqnarray}
Substituting the results from (\ref{ansatz}) and (\ref{smirnovresult}) into (\ref{completeset}),
\begin{eqnarray}
&&W_{\mu\nu}(x)_{a_0c_0e_0f_0}=\sum_{M=1}^{\infty}\int\left(\prod_{j=1}^{2M}\frac{d\theta_j}{4\pi}\right)e^{-ix\sum p}4\pi^2(4\pi)^{2M-2}\nonumber\\
&&\times(p_1+p_3+\dots+p_{2M-1}-p_2-\cdots-p_{2M})_\mu(p_1+\dots+p_{2M-1}-p_2-\cdots-p_{2M})_\nu\nonumber\\
&&\times(\delta_{a_0e_0}\delta_{c_0f_0}-\frac{1}{N}\delta_{a_0c_0}\delta_{e_0f_0})\prod_{j=1}^{2M-1}\left[\frac{1}{(\theta_j-\theta_{j+1})^2+\pi^2}\right]+\mathcal{O}\left(\frac{1}{N}\right).\nonumber
\end{eqnarray}

We have also found the form factors of the energy-momentum tensor operator, $T_{\mu\nu}(x)$. We start with a form-factor {\em Anzatz}, similar to (\ref{ansatz}). The energy-momentum tensor operator has two Lorentz indices and no $SU(N)$-color indices, while the Noether current has one Lorentz index and two color indices. The most general color- and Lorentz-invariant expression for the 2$M$-excitation form factor is 
\begin{eqnarray}
&&\langle 0\vert T_{\mu\nu} (x) \vert A,\theta_1,b_1,a_1;\dots;A,\theta_M,b_M,a_M;P,\theta_{M+1},a_{M+1},b_{M+1};\dots;P,\theta_{2M},a_{2M},b_{2M}\rangle\nonumber\\
&&=(p_1+\cdots +p_{M}-p_{M+1}-\cdots -p_{2M})_\mu(p_1+\cdots +p_{M}-p_{M+1}-\cdots -p_{2M})_\nu\nonumber\\
&&\times\frac{e^{-ix\cdot \sum p}}{N^{M-1}}\sum_{\sigma,\tau\in S_{M}} F_{\sigma\tau}(\theta_1,\dots,\theta_{2M}) \prod_{j=1}^M\delta_{a_j a_{\sigma(j)+M}}\prod_{k=1}^M\delta_{b_k b_{\tau(k)+M}}.\label{ansatzmomentum}
\end{eqnarray}
The Smirnov form-factor axioms are used to calculate the remaining functions $F_{\sigma\tau}(\{\theta\})$. This calculation is shown in more detail in Reference \cite{correlations}. The final result for the exact two point function is
\begin{eqnarray}
&&W^T_{\mu\nu\alpha\beta}(x)=\frac{1}{N^2}\langle0|T_{\mu\nu}(x)T_{\alpha\beta}(0)|0\rangle
=\sum_{M=1}^{\infty}\frac{\pi}{8}\int\left(\prod_{j=1}^{2M}d\theta_j\right)e^{-ix\sum p}\nonumber\\
&&\times(p_1+p_3+\cdots+p_{2M-1}-p_2-\cdots-p_{2M})_\mu(p_1+\cdots+p_{2M-1}-p_2-\cdots-p_{2M})_\nu\nonumber\\
&&\times(p_1+p_3+\cdots+p_{2M-1}-p_2-\cdots-p_{2M})_\alpha(p_1+\cdots+p_{2M-1}-p_2-\cdots-p_{2M})_\beta\nonumber\\
&&\times\frac{1}{[(\theta_1-\theta_{2M})^2+\pi^2]}\prod_{j=1}^{2M-1}\frac{1}{[(\theta_j-\theta_{j+1})^2+\pi^2]}+\mathcal{O}\left(\frac{1}{N}\right).\label{twopointmomentum}
\end{eqnarray}

\section{2+1 Yang-Mills as a Nearly Integrable Model}

We are interested in an anisotropic version of Yang-Mills theory, where two of the coordinates, $x^0
,{\rm and}\,\, x^1$, are rescaled to $ \lambda x^0, \,\lambda x^1$. The rest of the coordinates are unchanged. The gauge field components transform as $A_{0,1}\to(1/\lambda) A_{0,1}$. The $\lambda\to 0$ limit was first explored in \cite{verlinde}, \cite{mv}, in the context of hadron or heavy-ion collisions. This rescaling was studied in terms of an anisotropic renormalization group in References \cite{rescaling}. We are interested in this limit, because Yang-Mills theory in 2+1 and 3+1 dimensions has been shown to be integrable at $\lambda=0$ \cite{ym}.
 
 This rescaling can be visualized by starting with (2+1)-dimensional Yang-Mills theory regularized on a lattice with spacing $a$. The rescaling of coordinates amounts to taking $a$ to zero in the $x^0$ and $x^1$ directions. The system becomes an array of (1+1)-dimensional models coupled together to form a (2+1)-dimensional model. 
In the axial gauge, $A_1=0$, the rescaled Hamiltonian is $H=H_0+\lambda^2 H_1$, where 
\begin{eqnarray}
H_0=\sum_{x^2} H_{PCSM}(x^2)=\sum_{x^2} H_0 (x^2),\nonumber
\end{eqnarray}
and
\begin{eqnarray}
&&\lambda^2 H_1=\sum_{x^2}\lambda^2 H_1(x^2,x^2-a)=-\sum_{x^2} \int dx^1\int dy^1 \frac{\lambda^2}{4g_0^2a^2}\vert x^1-y^1\vert\nonumber\\
&&\times [j_0^L(x^1,x^2)-j_0^R(x^1,x^2-a)][j_0^L(y^1,x^2)-j_0^R(y^1,x^2-a)],\nonumber\\
\end{eqnarray}
so that there is a (1+1)-dimensional principal chiral sigma model Hamiltonian at each value of $x^2$. The coupling between two neighboring sigma models is given by $\lambda^2 H_1$. The $SU(N)$-group-valued field of the chiral model is defined from the remaining gauge field component, $U=e^{iaA_2}$.

Physical Yang-Mills states, $\Psi$, satisfy Gauss's Law, which after axial gauge fixing, and coordinate rescaling becomes
\begin{eqnarray}
\int dx^1[j_0^L(x^1,x^2)-j_0^R(x^1,x^2-a)]\Psi=0.\nonumber
\end{eqnarray}
This means that the particles of the principal chiral model particles form color singlets. The lowest energy excitation is one with a particle and an antiparticle in one of the sigma models, with both color indices contracted.

In the future we hope to examine corrections for non-zero $\lambda$, away from the integrable limit, in the context of form factor perturbation theory \cite{mussardo}. This involves computing matrix elements  $\langle \Psi^\prime \vert H_1 \vert \Psi\rangle$.  This is equivalent to evaluating Noether current correlation functions between the states of the principal chiral model. This is exactly what we have found at large $N$.

 The isotropic theory can be examined through the truncated spectrum approach. This was used by R.M. Konik and Y. Adamov to explore the 3-dimensional Ising model as an array of coupled 2-dimensional Ising chains \cite{konik}. One can discretize the spectrum of the (1+1)-dimensional models by putting them in a box of finite size. The physical states are then ordered by energy,  as $\vert 1\rangle,\,\vert 2\rangle,\dots,\,\vert n\rangle$, with energies $E_1 < E_2< \dots< E_n$, respectively, where $E_n$ is the truncation energy.

We can define a transfer-matrix operator that describes how the system evolves in the $x^2$ direction, as
\begin{eqnarray}
\hat{T}_{x^2-a,\,x^2}=e^{-\frac{1}{2} H_0(x^2-a)-\frac{1}{2}H_0(x^2)-\lambda^2H_1(x^2,\,x^2-a)}.\nonumber
\end{eqnarray}
In the truncated spectrum approach one can build a discrete, $n\times n$ matrix $T_{ij}=\langle i\vert \hat{T}_{x^2-a,\, x^2}\vert j\rangle$, using the set of states with energies $E_i,\,E_j\leq E_n$. The Yang-Mills partition function is 
\begin{eqnarray}
Z={\rm Tr} \,T^{N_2},\nonumber
\end{eqnarray}
where $N_2$ is the total number of sigma models (the size of the $x^2$ direction). The partition function can be computed by diagonalizing the matrix $T_{ij}$, which can be done numerically, or perturbatively in powers of $\lambda$. One can extract the bound state masses this way, and examine their dependence on the truncation energy $E_n$.

\end{document}